\documentclass[10pt,twocolumn,letterpaper]{article}

\usepackage{cvpr}              

\usepackage[dvipsnames]{xcolor}

\definecolor{cvprblue}{rgb}{0.21,0.49,0.74}
\usepackage[pagebackref,breaklinks,colorlinks,citecolor=cvprblue]{hyperref}
\usepackage{algorithm}
\usepackage{algpseudocode}
\usepackage{bigstrut}
\usepackage{multirow}
\usepackage{colortbl}
\usepackage{xcolor}
\def\paperID{14189} 
\def\confName{CVPR}
\def\confYear{2024}

\title{PolarRec: Radio Interferometric Data Reconstruction with Polar Coordinate Representation}

\author{Ruoqi Wang, Zhuoyang Chen, Jiayi Zhu\\
HKUST(GZ)\\
\and
Qiong Luo \\
HKUST(GZ) \& HKUST
\and
Feng Wang\\
GZHU
}

\begin{document}
\maketitle
\begin{abstract}
In radio astronomy, visibility data, which are measurements of wave signals from radio telescopes, are transformed into images for observation of distant celestial objects. However, these resultant images usually contain both real sources and artifacts, due to signal sparsity and other factors. One way to obtain cleaner images is to reconstruct samples into dense forms before imaging. Unfortunately, existing reconstruction methods often miss some components of visibility in frequency domain, so blurred object edges and persistent artifacts remain in the images. Furthermore, the computation overhead is high on irregular visibility samples due to the data skew. To address these problems, we propose PolarRec, a transformer-encoder-conditioned reconstruction pipeline with visibility samples converted into the polar coordinate representation. This representation matches the way in which radio telescopes observe a celestial area as the Earth rotates. As a result, visibility samples distribute in the polar system more uniformly than in the Cartesian space. Therefore, we propose to use radial distance in the loss function, to help reconstruct complete visibility effectively. Also, we group visibility samples by their polar angles and propose a group-based encoding scheme to improve the efficiency. Our experiments demonstrate that PolarRec markedly improves imaging results by faithfully reconstructing all frequency components in the visibility domain while significantly reducing the computation cost in visibility data encoding. We believe this high-quality and high-efficiency imaging of PolarRec will better facilitate astronomers to conduct their research.
\end{abstract}    
\section{Introduction}
In radio astronomy, \emph{visibility} refers to radio signal data from celestial objects, obtained by radio telescopes.  These data are represented as complex values in the \emph{uv-plane}, a geometric plane defined for interferometric observations. Visibility data are subsequently converted into images through \emph{imaging} for further analysis. However, these images, known as \emph{dirty images}, are often dominated by artifacts \cite{schmidt2022deep}. This phenomenon is due to limitations in telescope configurations, under which not the entire uv-plane is sampled.  Therefore, visibility data normally have to be reconstructed before being utilized in scientific analysis. In this paper, we propose a visibility reconstruction method, aiming to reconstruct the real sky by recovering all visibility components in the uv-plane effectively and efficiently.

\begin{figure}
  \centering
  \begin{subfigure}{0.9\linewidth}
    \includegraphics[width=7.2cm]{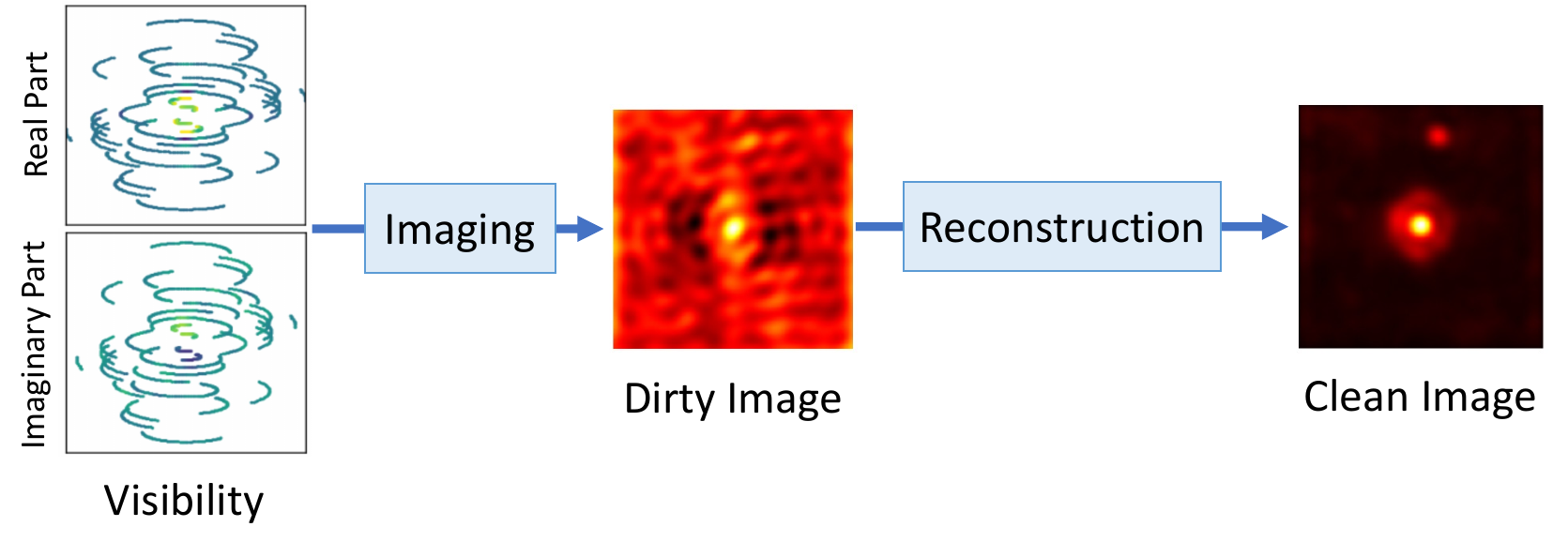} 
    \caption{Traditional method: imaging followed by reconstruction}
  \end{subfigure}
  \\
  \begin{subfigure}{0.9\linewidth}
    \includegraphics[width=7.2cm]{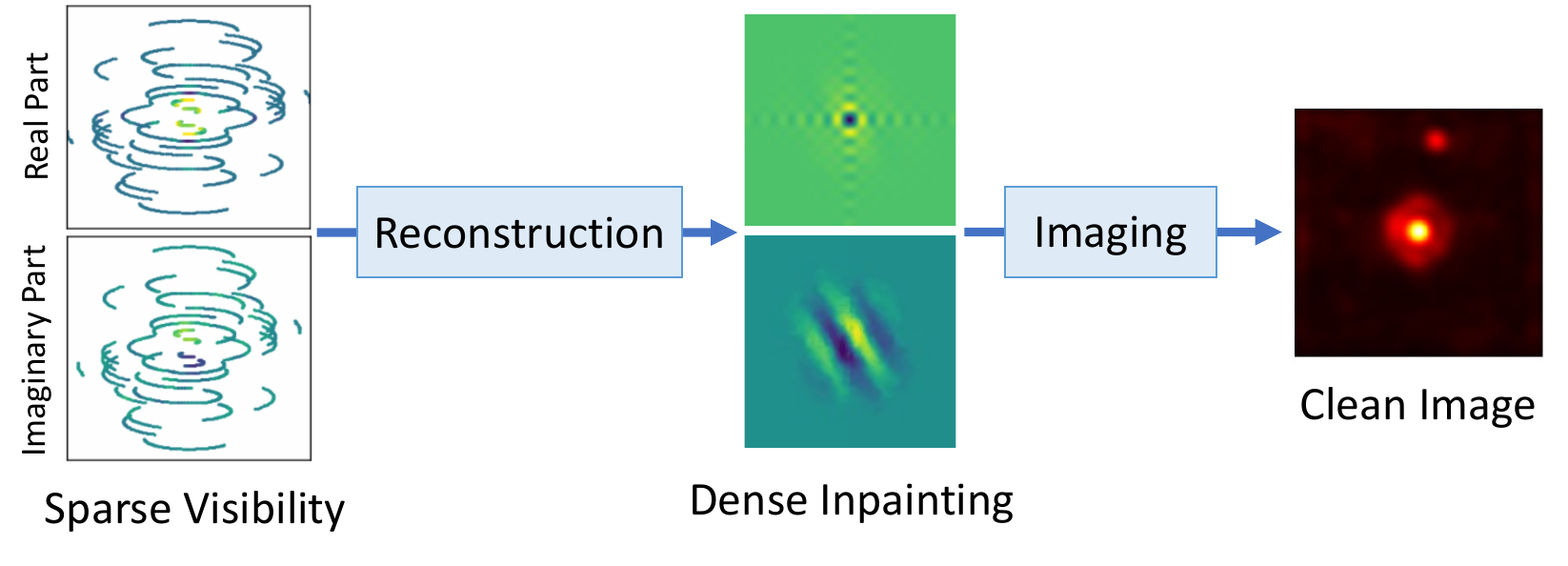}
    \caption{Recent method: reconstruction followed by imaging.}
  \end{subfigure}
  \caption{Two visibility data processing flows.}
  \label{flow}
\end{figure}

Traditional methods first transfer the sparse visibility data into dirty images through the imaging process and then reconstruct the dirty images to \emph{clean images} \cite{hogbom1974aperture,ables1974maximum,bouman2016computational,sun2021deep,connor2022deep,WangC0W23}. The process is shown in Figure \ref{flow} (a). In contrast, some recent deep-learning-based studies \cite{schmidt2022deep,wu2022neural} have proposed to first do inpainting on the visibility data to reconstruct the sparse samples to dense coverage and then perform imaging to obtain the clean image, as shown in Figure \ref{flow} (b), showing better performance. In order to produce a high-fidelity sharp image, it is crucial to densely sample the full visibility domain \cite{wu2022neural}. 
In this paper, we adopt the reconstruction-and-imaging processing flow.

Existing methods for reconstructing visibility data face challenges in both effectiveness and efficiency.
Specifically, the first challenge is to effectively capture visibility components within the uv-plane. For example, Radionets \cite{schmidt2022deep} and U-Nets \cite{ronneberger2015u} use a convolutional neural network that is based on pixels and grids, resulting in discontinuities in reconstructed visibility maps. In comparison, Wu et al. \cite{wu2022neural} use neural fields to address the continuity problem, but their method can only restore the low-frequency part of visibility (located near the center of the uv-plane), largely missing the high-frequency portion (found far from the uv-plane center). Such discontinuity and incompleteness in the visibility domain cause blurred edges of observed objects, disappearance of faint astronomical sources, and persistence of artifacts, in resultant images. 

The second major issue is the inefficiency of current strategies. For instance, Radionets \cite{schmidt2022deep} encode visibility data together with an excessive amount of zero-inpainting, causing a large amount of unnecessary computation. Wu et al. \cite{wu2022neural} embed each visibility sample point as a token in their Transformer encoder \cite{vaswani2017attention} to attend to each other, which is of quadratic computation cost in the number of sample points \cite{dosovitskiy2021image}, too high for real applications. These limitations force a compromise between computational cost and quality of reconstruction in current methods, severely limiting the application of deep learning techniques in radio astronomy imaging and reconstruction.

\begin{figure}
  \centering
  \begin{subfigure}{0.48\linewidth}
  \centering
    \includegraphics[width=3.2cm, height=2.9cm]{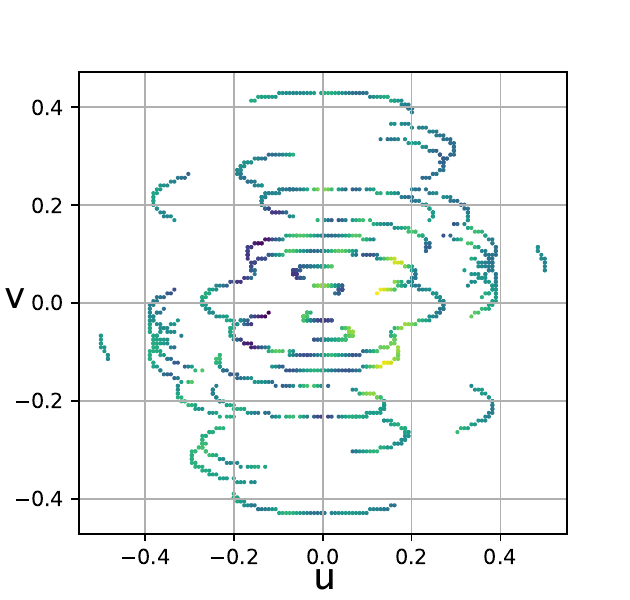} 
    \caption{Cartesian Representation.}
  \end{subfigure}
  \begin{subfigure}{0.48\linewidth}
  \centering
    \includegraphics[width=3.2cm]{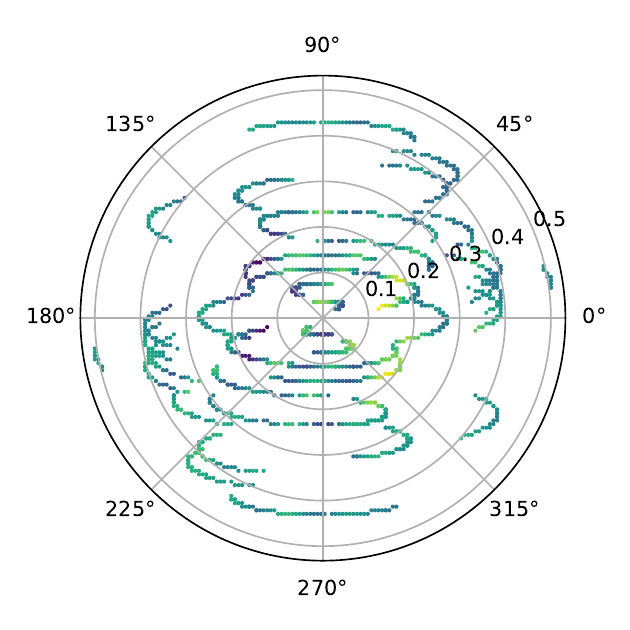}
    \caption{Polar Representation.}
  \end{subfigure}
  \caption{Visibility data samples with different representations.}
  \label{representations}
\end{figure}

To address these challenges, we explore the observational nature of radio telescopes and then propose PolarRec, which leverages the polar representation of sample points to enhance the reconstruction quality and reduce the cost. Our key observation is that visibility samples are obtained by telescopes as the Earth rotates, and that the high- and low-frequency signals are distributed on the uv-plane in accordance with their distances from the center. As a result, Cartesian-based approaches cannot represent the visibility points evenly \cite{nie2023partner} (see Figure \ref{representations}). In contrast, the polar-based representation has shown promising properties in vision tasks including instance segmentation and object detection \cite{nie2023partner,xie2020polarmask,xie2021polarmask++,chen2022polar}. Therefore, we propose to adopt the polar representation, using the radial coordinate to associate with the frequency information in the loss function and the angular coordinate to group sample points.  

Specifically, we first design a loss function for visibility data with a weighting scheme based on the radial coordinates. Our weighted loss function, Radial Visibility Loss, associates visibility data with radial coordinates in the frequency domain, enabling effective reconstruction of visibility data, on both high- and low-frequency components. In comparison, existing work recovers mainly low-frequency components. Consequently, our method produces sharper, more detailed imaging results. Furthermore, by introducing group-level encoding of visibility points according to angular coordinates, our method is more efficient than visibility-point-level encoding. This group encoding improves computation efficiency, making the use of Transformer encoders for visibility encoding more practical and scalable.

In summary, our main contributions are as follows:
\begin{itemize}
    \item We propose PolarRec to seamlessly integrate visibility data with a Transformer-encoded reconstruction pipeline by using polar coordinate representation, addressing two fundamental challenges in visibility reconstruction.
    \item We introduce the Radial Visibility Loss (RVL), which incorporates the radial coordinates of visibility data, adeptly capturing both low- and high-frequency visibility components.
    \item As the first to study the granularity of visibility encoding, we innovatively utilize angular coordinates to group sampled points for Transformer encoding, markedly enhancing computational efficiency.
\end{itemize}

We have experimentally evaluated our proposed PolarRec on public datasets of four galaxy morphologies, including an overall comparison with other state-of-the-art methods, ablation studies, and tests on group sizes and grouping techniques. The experimental results demonstrate the effectiveness and efficiency of our method. 
\section{Background and Related Work}
\label{Related Work}
\subsection{Very Long Baseline Interferometry (VLBI)}
In radio astronomy, using radio interference signals to image distant astronomical sources requires telescopes of very large aperture \cite{bouman2018reconstructing}, because the angular resolution of a telescope is inversely proportional to the diameter. A major observation technique is the Very Long Baseline Interferometry (VLBI), which uses multiple radio telescopes spreading over the globe to form a virtual Earth-sized telescope. The radio waves from astronomical sources are recorded separately at individual telescopes. Then, these signals are cross-correlated for all pairs of antennas at a central location, generating \emph{visibility} data. A VLBI observation is typically performed for hours to measure as many points in the uv-plane as possible. However, the measurement results remain sparse due to the limited number of antennas \cite{thompson2017interferometry,bouman2018reconstructing}. Consequently, sparse-to-dense reconstruction on visibility data is necessary to improve the imaging quality for the following analysis. 

\subsection{Interferometric Imaging}
Visibility data, represented as complex values, is the result of a Fourier transform of the sky's brightness distribution \cite{liu2022efficient}. \emph{Imaging} converts visibility data into images, which can be analyzed to provide insights about the observed celestial bodies \cite{thompson2017interferometry}. In the imaging process, an inverse Fourier transform maps the $(u, v)$ coordinates from the Fourier domain to $(l, m)$ coordinates in the image domain \cite{wu2022neural}. The transformation can be described as:
\begin{eqnarray}
I(l, m) & = & \int_{u} \int_{v} e^{2 \pi i(u l+v m)} V(u, v) d u d v.
\end{eqnarray}
In this equation, $V(u, v)$ is the visibility data in Fourier space, and $I(l, m)$ represents the intensity distribution in the image domain.

\subsection{Radio Interferometric Data Reconstruction}

Radio interferometric data reconstruction is vital for converting sparse visibility data into clear, artifact-free images. However, previous methods in this domain \cite{hogbom1974aperture,ables1974maximum,bouman2016computational,sun2021deep,connor2022deep,WangC0W23,wu2022neural,schmidt2022deep} have been limited by either their inability to capture continuous and complete frequency components or computational inefficiencies. To the best of our knowledge, no previous studies have managed to reconstruct all components in visibility data or increase the granularity of visibility encoding for efficiency.
\begin{figure}
\centerline{\includegraphics[height=4.1cm]{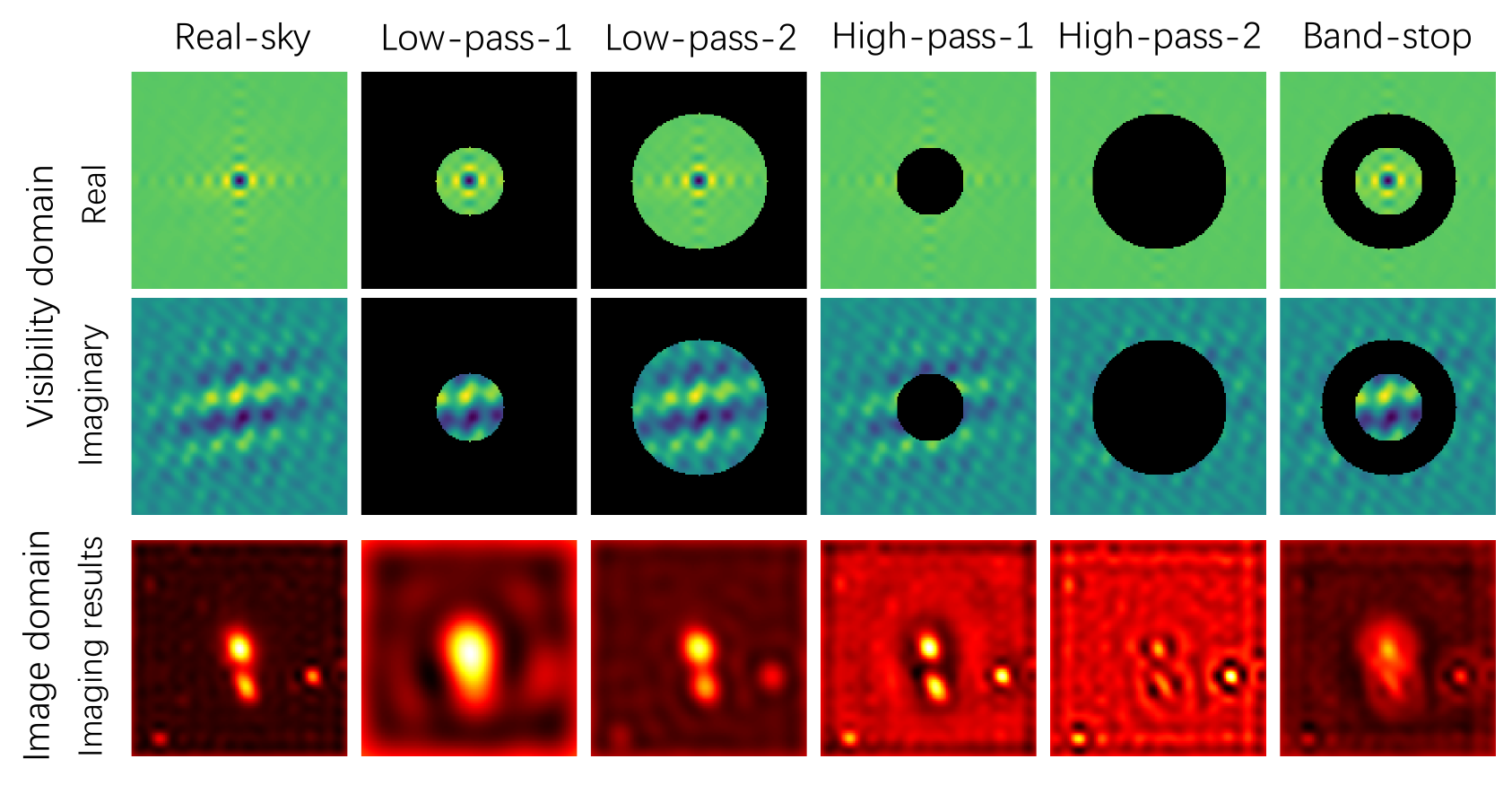}}
\caption{Effects of band limiting.}
\label{bandilmiting}
\end{figure}

\section{Our Method}
\label{Method}
\begin{figure*}[t]
\centerline{\includegraphics[height=5.6cm]{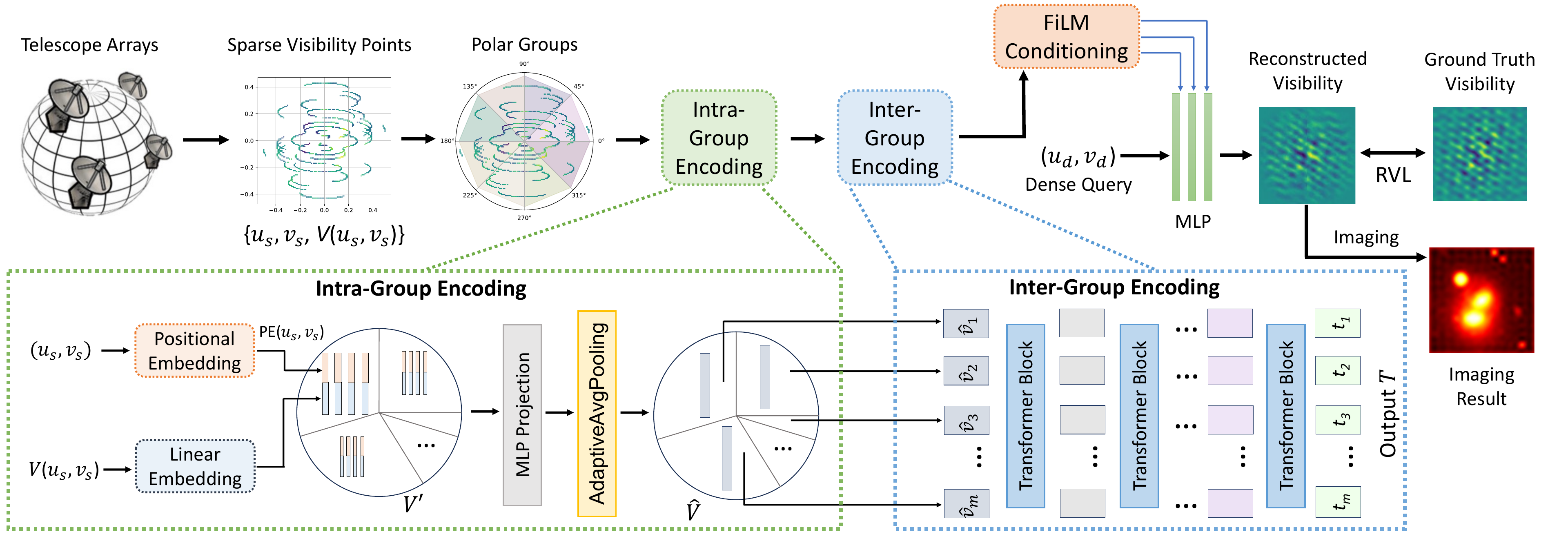}}
\caption{An overview of our method. Sparse visibility data ${V}\left(u_{s}, v_{s}\right)$ are grouped by the angular coordinate and passed through two encoding layers: intra-group encoding to generate group tokens and inter-group encoding by a Transformer encoder. The encoded output $T$ then conditions the predicted visibility generation in the neural field. The final output is compared with the ground truth to compute the Radial Visibility Loss.}
\label{Overview}
\end{figure*}
In this section, we first investigate the relation between visibility components and imaging results. After that, we present PolarRec, which adopts polar coordinate representation in visibility reconstruction. In PolarRec, we design Radial Visibility Loss to incorporate a weighting scheme based on the radial coordinate within the uv-plane. Moreover, we group the visibility points by their angular coordinates and then extract grouping tokens for the subsequent Transformer encoding. An overview of our method is presented in Figure \ref{Overview}. 

\subsection{Imaging Results of Visibility}

In radio interferometry, visibility data plays a vital role in generating high-fidelity images of celestial objects. The imaging process is the inverse Fourier transformation of visibility data to construct the brightness distribution of the observed sky. Consequently, missing or incomplete visibility components can significantly degrade the imaging result, leading to artifacts and loss of crucial information about the object's structure. Therefore, we first investigate the impact of missing visibility components on the final imaging output. 

We explore the impact of visibility components in various frequency regions by applying standard band-limiting operations \cite{jiang2021focal} and analyzing the effects of individual components on the imaging results. As shown in Figure \ref{bandilmiting}, loss of high frequency data due to a low-pass filter (Column low-pass-1 and low-pass-2) results in blur and artifacts and causes the vanishing of weak sources in the imaging results. Comparing low-pass-1 and low-pass-2, we can see that the larger the radius of the retained portion in the frequency domain, the clearer the resulting image is, and the fewer artifacts there are. In comparison, when low frequency data are absent due to a high-pass filter (Column high-pass-1 and high-pass-2), the overall quality of the image declines, but the object edges are clear, and dim or small sources around the main observed object are retained. Last, the band-and-stop filter (Column Band-stop) also causes artifacts and blur in imaging results.

In summary, different frequency regions cause distinct imaging effects. This observation suggests that recovering all missing visibility components could enhance the quality of imaging results. More specifically, if a resultant image has blurred edges or misses dim light sources, it is probably due to poor reconstruction of the high-frequency components of the visibility data.

\subsection{Polar Coordinate Representation}
Our utilization of the polar coordinate representation is to represent the visibility more uniformly. It is because visibility sampling is based on Earth's rotation, and high- and low-frequency visibility components are distributed on the uv-plane according to their distances from the origin of the plane. 

In the uv-plane, we convert $(u, v)$ coordinates to polar coordinates, denoted as $(r(u, v), \theta(u, v))$:

\begin{equation}
r(u, v) = \sqrt{u^2 + v^2}
\end{equation}

\begin{equation}
\theta(u, v) = \text{arctan2}(v, u)
\end{equation}

Where $r(u, v)$ represents the radial distance from the origin, and $\theta(u, v)$ represents the angle of the vector from the positive $u$-axis.

\subsection{Radial Visibility Loss}

We propose Radial Visibility Loss (RVL), incorporating the radial coordinates of visibility data to capture both low- and high-frequency visibility.

To compute RVL, we first compute the weight matrix $w_1(u, v)$ to down-weight easy visibility components (components whose predicted values are close to the ground truth) based on Focal Frequency Loss (FFL) \cite{jiang2021focal}:
\begin{equation}
w_1(u, v) = |V_r(u, v) - V_p(u, v)|^\alpha
\end{equation}
where $V_r(u, v) = A_r + iB_r $ and $ V_p(u, v) = A_p + iB_p$ are the ground truth and predicted visibilities in complex form respectively, and $\alpha$ is a scaling factor.

We then introduce an additional weight $w_2(u, v)$, computed from $r(u, v)$, to make our model pay more attention to high-frequency components of the visibility during reconstruction. The weight term $w_2(u, v)$ is calculated as follows:

\begin{equation}
w_2(u, v) = \left(\frac{r(u, v)}{\max(r(u, v))} + 1\right)^\beta 
\end{equation}
Where $\frac{r(u, v)}{\max(r(u, v))}$ normalizes the radial coordinate, ensuring the weight is more significant for points farther from the center, corresponding to higher frequencies in the visibility data. Adding 1 prevents any weights from becoming zero, maintaining the influence of all visibility components during the learning process. $\beta$ is a scaling factor.

The final weight $w(u, v)$ is computed as:

\begin{equation}
w(u, v) = \left(\frac{r(u, v)}{\max(r(u, v))} + 1\right)^\beta |V_r(u, v) - V_p(u, v)|^\alpha
\end{equation}

\begin{figure}[b]
\centerline{\includegraphics[height=3.6cm]{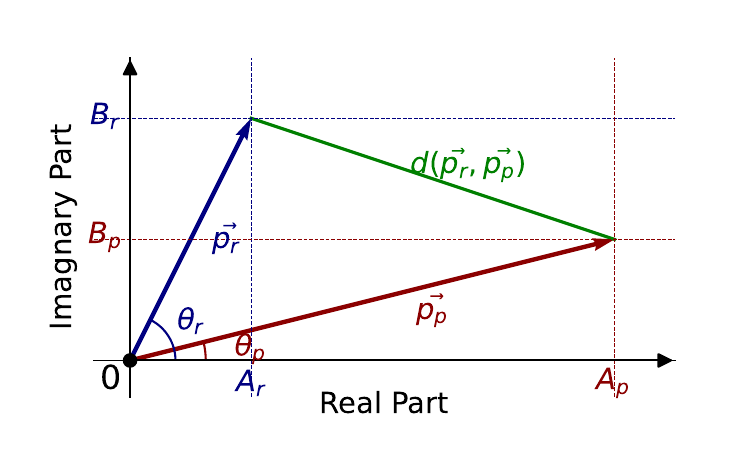}}
\caption{Example of visibility distance.}
\label{visdis}
\end{figure}

A thorough visibility distance measure must consider both amplitude and phase as they hold distinct aspects of imaging information \cite{jiang2021focal}. The complex visibility value can be represented as a vector within a two-dimensional plane. Specifically, we map ground truth visibility \( V_r(u, v) = A_r + iB_r \) and the corresponding predicted value \( V_p(u, v) = A_p + iB_p \) to vectors \( \vec{p_r} \) and \( \vec{p_p} \) (as shown in Figure \ref{visdis}). According to the definition of 2D discrete Fourier transform, the vector magnitude relates to the amplitudes $|\vec{p_r}|$ and $|\vec{p_p}|$, and the angles $\theta_r$ and $\theta_p$ reflect the phase. Consequently, following FFL \cite{jiang2021focal}, we define the single-point visibility distance considering both magnitude and direction as: 
\begin{equation}
d(\vec{p_r}, \vec{p_p}) = \| \vec{p_r} - \vec{p_p} \|^2_2 = |V_r(u, v) - V_p(u, v)|^2.
\end{equation}
The final visibility distance is computed as the average of these individual distances over all visibility points:
\begin{equation}
d(V_r, V_p) = \frac{1}{MN} \sum_{u=0}^{M-1} \sum_{v=0}^{N-1} |V_r(u, v) - V_p(u, v)|^2.
\end{equation}

The final expression for the Radial Visibility Loss is derived by executing the Hadamard product between the weight matrix $w(u, v)=w_1(u, v)w_2(u, v)$ and the visibility distance matrix $d(V_r, V_p)$, as follows:

\begin{equation}
\text{RVL} = \frac{1}{MN} \sum_{u=0}^{M-1} \sum_{v=0}^{N-1} w(u, v)|V_r(u, v) - V_p(u, v)|^2
\end{equation}

\subsection{Encoding by Angular-Coordinate Groups}
Using Transformers for encoding visibility samples causes quadratic computation cost in the number of sample points \cite{dosovitskiy2021image}. Such high costs constrain practical applications, especially as the number of visibility points in each image expands with more telescope arrays or extended observation duration. To solve the problem, we propose a group-based encoding scheme to improve the efficiency.

As shown on the left in Figure \ref{Overview}, the sparsely sampled visibility data is in the form of $\left\{u_{s}, v_{s}, {V}\left(u_{s}, v_{s}\right)\right\}$, where $(u_{s}, v_{s})$ are the coordinates at which a measurement is sampled and ${V}\left(u_{s}, v_{s}\right)$ is the complex value of the sample point. 

First, we divide all sample points into $m$ groups according to their angular coordinate $\theta(u_s, v_s)$.
To integrate each visibility value and its corresponding position, we encode each sample individually using positional embedding ($\text{PE}(u_s,v_s)$ in Figure \ref{Overview}). Specifically, we encode the positional information of a sample point using Random Fourier Embedding \cite{tancik2020fourier}. After that, the positional embedding $\text{PE}(u_s,v_s)$ and the embedding of complex value of visibility ${V}\left(u_{s}, v_{s}\right)$ are concatenated to form visibility tokens $V^{\prime}$. Denote the sparse visibility tokens as $V^{\prime}=\left[v^{\prime}_{1} ; v^{\prime}_{2} ; v^{\prime}_{3} ; \ldots v^{\prime}_n\right]$, where $V^{\prime} \in \mathbb{R}^{n \times d}$, $n$ is the number of sample measurement points and $d$ is the number of dimensions of visibility tokens.

Then, we apply a Multi-Layer Perceptron (MLP) mapping layer and averaging aggregating for intra-group encoding. The encoding result is  $\hat{V}=\left[\hat{v}_{1} ; \hat{v}_{2} ; \hat{v}_{3}; \ldots \hat{v}_m\right]$, where $\hat{V} \in \mathbb{R}^{m \times d}$.  For each group $i$, we compute the group token $\hat{v}_i$ as follows:

\begin{eqnarray}
\hat{v}_i = \text{Avg}\left(\text{MLP}(v_j)\right), \text{  }i=1 \text{ to } m, \text{  }v_j\in \text{ group } i.
\end{eqnarray}

In the implementation of intra-group encoding, we sort the tokens $V^{\prime}$ of sample points according to their angular coordinates. After mapping these tokens through MLP, we use adaptive average pooling to compute the group tokens $\hat{V}$. For data collected with the same telescope configuration, the process of sorting based on angular coordinates needs to be performed only once, as the locations of sample points remain unchanged. The intra-group encoding process is as shown in Algorithm \ref{al:groupToken}.

\begin{algorithm}
\caption{Intra-Group Encoding}
\label{al:groupToken}
\small
\begin{algorithmic}[1]
    \For{each $u_{si}, v_{si}, {V}(u_{si}, v_{si})$ in set of sampled points}
         \State $\text{pe}_i \gets \text{PE}(u_{si}, v_{si})$
         \State $\text{le}_i \gets \text{LinearEmbedding}({V}(u_{si}, v_{si}))$
        \State $v_i^{\prime} \gets \text{concat}(\text{pe}_i, \text{le}_i)$
        \State $\theta_i \gets \text{arctan2}(u_{si}, v_{si})$
    \EndFor
    \State $V^{\prime} \gets \left[v^{\prime}_{1} ; v^{\prime}_{2} ; \ldots ; v^{\prime}_n\right], \Theta \gets \left[\theta_{1} ; \theta_{2} ; \ldots ; \theta_n\right]$ 

    \If{\textit{sorted\_indices} not precomputed}
        \State $\textit{sorted\_indices} \gets \text{argsort}(V^{\prime}, \Theta)$
    \EndIf

    \State $V^{\prime} \gets V^{\prime}[\textit{sorted\_indices}]$

    \State $V^{\text{MLP}} \gets \text{MLP}(V^{\prime\prime})$

    \State $\hat{V} \gets \text{AdaptiveAvgPooling}(V^{\text{MLP}}, m)$

    \State \textbf{return} $\hat{V}$
\end{algorithmic}
\end{algorithm}

Finally, the group tokens $\hat{V}$ go through inter-group encoding by a Transformer encoder. We base our encoder design on Transformer structures similar to prior work \cite{vaswani2017attention,dosovitskiy2021image,wu2022neural}. The input group tokens are then transformed into latent tokens by multi-headed self-attention layers.

\subsection{Neural Field Conditioning}

Our method follows the conditional neural field pipeline proposed by Wu et al. \cite{wu2022neural}.
Given the sparsely sampled visibility ${V}(u_s, v_s)$, our objective is to determine a neural field $\Phi$, fulfilling a constraint set by the function $F$:
\begin{equation}
F(\Phi(u_s, v_s), {V}(u_s, v_s)) = 0.
\end{equation}

We approximate this implicit function $\Phi(u, v)$ with an MLP of $l$ layers parameterized by weights $\Theta_m$. 

We use the output tokens of inter-group encoding $T=\left[t_{1} ; t_{2} ; t_{3}; \ldots t_l\right]$ to extend the neural field with a learning-based prior, with each token corresponding to an MLP layer. Using the FiLM conditioning \cite{perez2018film}, the output tokens modulate the $i$th layer's activation $\mathbf{x_i}$ by:
\begin{equation}
\operatorname{FiLM}(\mathbf{x_i}) = \gamma(t_i) \odot \mathbf{x_i}+\beta(t_i), i \in 1  \text{ to } l,
\end{equation}
where $\gamma$ and $\beta$ are simple affine layers with non-linearities and $\odot$ signifies a Hadamard product \cite{horn1990hadamard}.

The MLP parameters $\Theta_{m}$ and the encoder parameters $\Theta_{e}$ are jointly optimized during training:
\begin{equation}
\begin{split}
\min _{\Theta_{m}, \Theta_{e}} \text{RVL} \left(\Phi\left(u_d, v_d ;\left\{T\right\} ; \Theta_{m}\right), V_{\mathrm{gt}}(u_d, v_d)\right) ,
\\
\text{with } \left\{T\right\} = \Psi\left(\left\{u_s, v_s, V(u_s, v_s)\right\}; \Theta_e\right),
\end{split}
\end{equation}
where $(u_{d}, v_{d})$ are the dense coordinates in visibility plane and ${V_{gt}}\left(u_{d}, v_{d}\right)$ is the ground truth of visibility inpainting.

\section{Experiments}
\label{Experiments}

In this section, we conduct a comprehensive evaluation of our method in comparison with several classic and recent state-of-the-art methods to demonstrate the overall improvement achieved by our approach. We also design experiments to explore the effects of different grouping methods and group sizes on the reconstruction results. In addition, we conduct ablation experiments to study the effects of individual weighting techniques in RVL.

\subsection{Experimental Setup}

\textbf{Platform}. We conduct all experiments on a server with two AMD EPYC 7302 CPUs, 128GB main memory, and eight Nvidia RTX 3090 GPUs each with 24GB device memory. The server is equipped with an NVME 2TB SSD and four 1TB SATA hard disks. The operating system is Ubuntu 20.04. Our model is implemented in PyTorch 1.8.1 \cite{paszke2019pytorch}.

\textbf{Datasets}. In the experiments, we evaluate our method on different public datasets of four kinds of distinct galaxy morphologies \cite{Galaxy10, ciprijanovic2022semi}: Merging Galaxies (MG), In-between Round Smooth Galaxies (IRSG), Unbarred Tight Spiral Galaxies (UTSG), and Edge-on Galaxies with Bulge (EGB). These data are derived from sources under the DESI Legacy Imaging Surveys \cite{dey2019overview}, integrating contributions from the Beijing-Arizona Sky Survey (BASS) \cite{zou2017project}, the DECam Legacy Survey (DECaLS) \cite{blum2016decam}, and the Mayall z-band Legacy Survey \cite{silva2016mayall}. The visibility data are generated from these images using the eht-imaging toolkit \cite{chael2019ehtim,chael2018interferometric}, denoted by $\left\{u_{s}, v_{s}, {V}\left(u_{s}, v_{s}\right)\right\}$. The parameters for observation are adjusted to mirror an 8-telescope Event Horizon Telescope (EHT) setup \cite{wu2022neural}, with the EHT being one of the most prominent arrays leveraging VLBI techniques. 


\textbf{Methods under Comparison}.
We compare our method with three other methods for radio interferometry reconstruction, including the classic method CLEAN \cite{hogbom1974aperture}, which is for dirty image reconstruction, and two latest deep learning-based approaches for visibility data reconstruction -- Radionets \cite{schmidt2022deep} and Neural Interferometry \cite{wu2022neural}. We use the original code of these methods and follow the parameter setting in the original code for the best performance. All these methods are implemented on PyTorch. In addition, we test the magnetic resonance imaging (MRI) reconstruction U-Net \cite{xie2022measurement, ronneberger2015u} to reconstruct visibility data as supplementary baselines.

\begin{table*}[!t]
    \centering
    \scriptsize
    \caption{Overall performance comparison (mean and standard deviation) on different datasets. All comparison methods have significant difference with our method ($p < 0.01$).}
    \label{table_1}
    \renewcommand{\arraystretch}{1.5} 
    \setlength\tabcolsep{1.0pt}
    \resizebox{\textwidth}{!}{
    \begin{tabular}{r|cc|ccc|ccc|ccc|ccc}
        \hline
        \multirow{2}*{\textbf{Models}}&\multirow{2}*{\textbf{Img}}&\multirow{2}*{\textbf{Vis}}&\multicolumn{3}{c}{\textbf{MG}}\vline&\multicolumn{3}{c}{\textbf{IRSG}}\vline&\multicolumn{3}{c}{\textbf{UTSG}}\vline&\multicolumn{3}{c}{\textbf{EGB}}\\
        &&&$LFD\downarrow$&$PSNR\uparrow$&$SSIM\uparrow$&$LFD\downarrow$&$PSNR\uparrow$&$SSIM\uparrow$&$LFD\downarrow$&$PSNR\uparrow$&$SSIM\uparrow$&$LFD\downarrow$&$PSNR\uparrow$&$SSIM\uparrow$\\
        \hline
        \textbf{Dirty}&&&$N/A$&10.453 \tiny{(1.256)}&0.680 \tiny{(0.055)}&$N/A$&10.875 \tiny{(1.293)}&0.711 \tiny{(0.050)}&$N/A$&10.655 \tiny{(1.092)}&0.690 \tiny{(0.047)}&$N/A$&10.158 \tiny{(1.216)}&0.674 \tiny{(0.055)}\\
        \textbf{CLEAN} \cite{hogbom1974aperture}&\checkmark&&$N/A$&18.708 \tiny{(2.336)}&0.767 \tiny{(0.031)}&$N/A$&20.974 \tiny{(2.192)}&0.795 \tiny{(0.025)}&$N/A$&17.200 \tiny{(2.465)}&0.750 \tiny{(0.035)}&$N/A$&20.175 \tiny{(2.411)}&0.779 \tiny{(0.033)}\\
        \textbf{U-Net} \cite{xie2022measurement}&&\checkmark&1.171 \tiny{(0.198)}&18.126 \tiny{(1.906)}&0.818 \tiny{(0.021)}&1.235 \tiny{(0.230)}&19.770 \tiny{(1.925)}&0.828 \tiny{(0.024)}&1.256 \tiny{(0.227)}&14.877 \tiny{(1.701)}&0.786 \tiny{(0.032)}&1.219 \tiny{(0.234)}&19.232 \tiny{(2.078)}&0.822 \tiny{(0.024)}\\
        \textbf{Radionets} \cite{schmidt2022deep}&&\checkmark&1.580 \tiny{(0.202)}&19.687 \tiny{(1.897)}&0.836 \tiny{(0.022)}&1.499 \tiny{(0.206)}&21.369 \tiny{(1.887)}&0.854 \tiny{(0.022)}&1.419 \tiny{(0.231)}&20.828 \tiny{(2.003)}&0.844 \tiny{(0.022)}&1.545 \tiny{(0.220)}&20.322 \tiny{(1.809)}&0.836 \tiny{(0.023)}\\
        \textbf{Neu\_Int} \cite{wu2022neural}&&\checkmark&1.229 \tiny{(0.282)}&20.560 \tiny{(2.287)}&0.875 \tiny{(0.028)}&0.916 \tiny{(0.302)}&24.099 \tiny{(2.760)}&0.898 \tiny{(0.027)}&1.038 \tiny{(0.313)}&21.323 \tiny{(2.303)}&0.880 \tiny{(0.027)}&1.146 \tiny{(0.343)}&21.276 \tiny{(2.675)}&0.879 \tiny{(0.031)}\\
        \rowcolor{gray!20}\textbf{PolarRec\tiny{(Ours)}}&&\checkmark&\textbf{0.924} \tiny{\textbf{(0.284)}}&\textbf{24.206} \tiny{\textbf{(2.547)}}&\textbf{0.889} \tiny{\textbf{(0.029)}}&\textbf{0.666} \tiny{\textbf{(0.238)}}&\textbf{26.540} \tiny{\textbf{(2.653)}}&\textbf{0.911} \tiny{\textbf{(0.025)}}&\textbf{0.707} \tiny{\textbf{(0.241)}}&\textbf{25.430} \tiny{\textbf{(2.565)}}&\textbf{0.905} \tiny{\textbf{(0.025)}}&\textbf{0.707} \tiny{\textbf{(0.241)}}&\textbf{25.880} \tiny{\textbf{(2.529)}}&\textbf{0.906} \tiny{\textbf{(0.025)}}\\
        \hline
    \end{tabular}
    }
    \vspace{-1.0em}
\end{table*}

\textbf{Evaluation Metrics}. 
To measure differences in frequency data, we use the Log Frequency Distance (LFD) \cite{jiang2021focal}, which is defined as follows:
{\small \begin{eqnarray}
\text{LFD} = \log \left[\frac{1}{M N}\left(\sum_{u = 0}^{M-1} \sum_{v = 0}^{N-1}\left|V_{r}(u, v)-V_{p}(u, v)\right|^{2}\right)+1\right]
\end{eqnarray}}
where $V_r(u, v)$ and $V_p(u, v)$ represent the real and predicted visibility respectively. A lower LFD is better.

To evaluate the quality of images produced from the reconstructed visibility, we employ two common metrics: Peak Signal-to-Noise Ratio (PSNR) and Structural Similarity Index Measure (SSIM). PSNR quantifies the overall image quality of the resultant images, and SSIM quantifies the perceptual similarity to the ground truth images. We compute these two metrics using the scikit-image package \cite{singh2019basics}, which follows the formulas presented by Hore et al. \cite{hore2010image}. A higher PSNR and SSIM is better.

To evaluate the efficiency, we use inference time and Floating Point Operations (FLOPs). 
The inference time in the experiments is tested on a single Nvidia RTX 3090. We compare time performance between our method and Neural Interferometry only, because only these two are transformer-encoder based.

\begin{figure}
\centerline{\includegraphics[width=8.6cm]{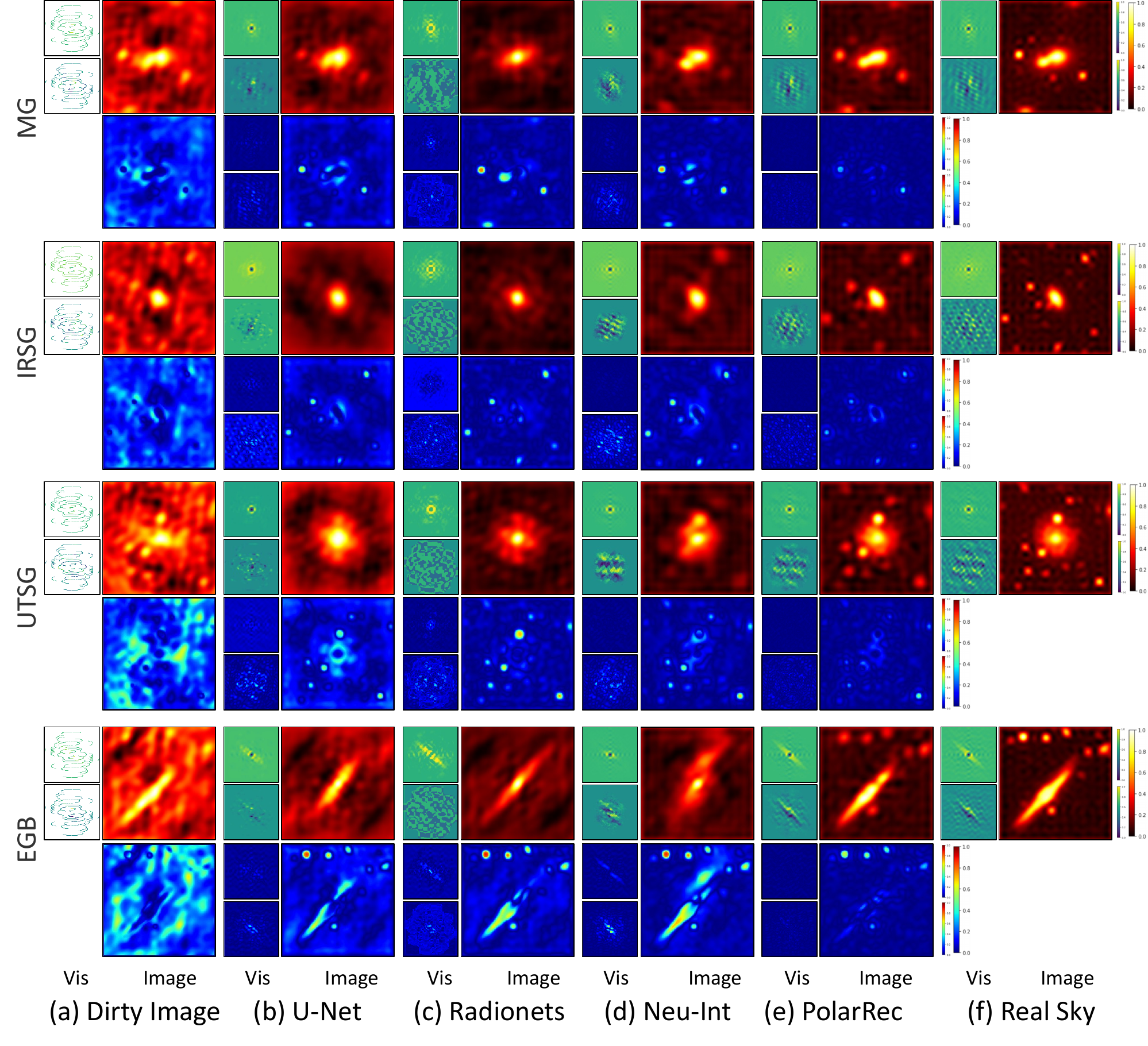}}
 \caption{Visual comparison of visibility and image. The reconstructed images, visibility data and their corresponding error maps are provided. The small panels to the left of each image are the corresponding visibility data, with the real part on top and the imaginary part at the bottom. }
\label{Visual}
\end{figure}

\subsection{Overall Comparison}
We calculate the LFD, PSNR, and SSIM values for all test images reconstructed using our method and other methods, presenting both mean values and standard deviations. These results are listed in Table \ref{table_1}.
The results show that PolarRec consistently outperforms the other methods in all three measures and four datasets, underscoring the effectiveness of our reconstruction method.

We also illustrate some representative reconstructed visibility and the corresponding images on the four datasets in Figure \ref{Visual}, including all deep-learning-based methods under comparison as well as the dirty images and the ground truth images of the real sky. Comparing the dirty images in Figure \ref{Visual} (a) and the ground truth in Figure \ref{Visual} (f), we find there are many artifacts and distortion of object structure in dirty images because of the sparsity of the visibility. 

As shown in Figure \ref{Visual} (b), reconstructions of the visibility by MRI reconstruction U-Net \cite{xie2022measurement} are dominated by the artifacts in resultant images. In Figure \ref{Visual} (c), Radionets \cite{schmidt2022deep} are able to reconstruct more visibility content than the U-Net. However, the reconstruction is discontinuous. Although Radionets reduce artifacts in the imaging results, it cannot distinguish between separate sources that are in close proximity. Furthermore, many faint astronomical sources are missing in the reconstructed images. In contrast, as shown in Figure \ref{Visual} (d), Neural Interferometry \cite{wu2022neural} (denoted Neu-Int) can continuously and realistically reconstruct the low-frequency components of the visibility, but misses much information from the high-frequency components, leading to a loss of details in the reconstruction. 
 
The results of PolarRec (Figure \ref{Visual} (e)) show that our method can effectively reconstruct more complete and continuous visibility data than others. The resultant imaging results not only eliminate artifacts but also restore the true structure of astronomical sources while preserving details and small faint sources.

\subsection{Computational Cost Comparison}
To demonstrate the effectiveness of the group-granularity encoding in PolarRec, we conduct a comparative analysis against the Neural Interferometry \cite{wu2022neural} method with point-granularity encoding due to both methods are built upon the Transformer architecture. We fix the number of visibility points as 1660 and the size of the reconstructed visibility map as 128 $\times$ 128, and compare the computational efficiency of the encoders with batch size from 4 to 32. The comparison focuses on two main metrics: the number of Floating Point Operations (GFLOPs) and inference time (Latency (ms)).

Table~\ref{tab:efficiency} shows the results of this comparison. The latency of group granularity encoding in PolarRec is significantly lower than point granularity encoding in Neural Interferometry \cite{wu2022neural}. Moreover, as the batch size increases, the rise in latency for PolarRec is much smaller compared to Neural Interferometry \cite{wu2022neural}. These results indicate that PolarRec has greater advantages in handling large batches of data, which has important implications for the processing and analysis of massive astronomical observational data.

\begin{table}[h]
\centering
\scriptsize
\caption{Comparative Analysis of Computational Efficiency}
\label{tab:efficiency}
\begin{tabular}{c|c|c|c}

\hline
\textbf{Method} & \textbf{BatchSize} & \textbf{Latency (ms) } & \textbf{GFLOPs} \\ \hline
                     & 4  & 29.53 \tiny{(3.33)} & 44.02 \\  
            Neural           & 8  & 59.02 \tiny{(3.19)} & 88.04 \\ 
       Interferometry            & 16 & 115.62 \tiny{(7.30)} & 176.08 \\ 
                                       & 32 & 232.04 \tiny{(5.64)} & 352.16 \\ \hline
                                     & 4  & 3.72 \tiny{(1.84)} & 3.05 \\ 
             PolarRec               & 8  & 3.89 \tiny{(3.23)} & 6.10 \\ 
\tiny{Group Size = 32}            & 16 & 4.20 \tiny{(2.00)} & 12.20 \\  
                                       & 32 & 5.37 \tiny{(3.05)} & 24.40 \\ \hline
                                & 4  & 3.47 \tiny{(0.10)} & 2.39 \\ 
             PolarRec             & 8  & 3.90 \tiny{(2.94)} & 4.78 \\ 
    \tiny{Group Size = 64}           & 16 & 4.02 \tiny{(2.11)} & 9.57 \\ 
                                       & 32 & 5.11 \tiny{(1.91)} & 19.14 \\ \hline
\end{tabular}
\end{table}

\subsection{Effect of Group Size}
In this experiment, we vary the group size in number of visibility points and investigate its impact on the reconstruction results.  We also record the inference latency of the encoding process, and use Floating Point Operations (FLOPs) to measure the computation cost. LFD, PSNR and SSIM are used to assess the reconstruction quality.

\begin{figure}[h]
\centerline{\includegraphics[height=6cm]{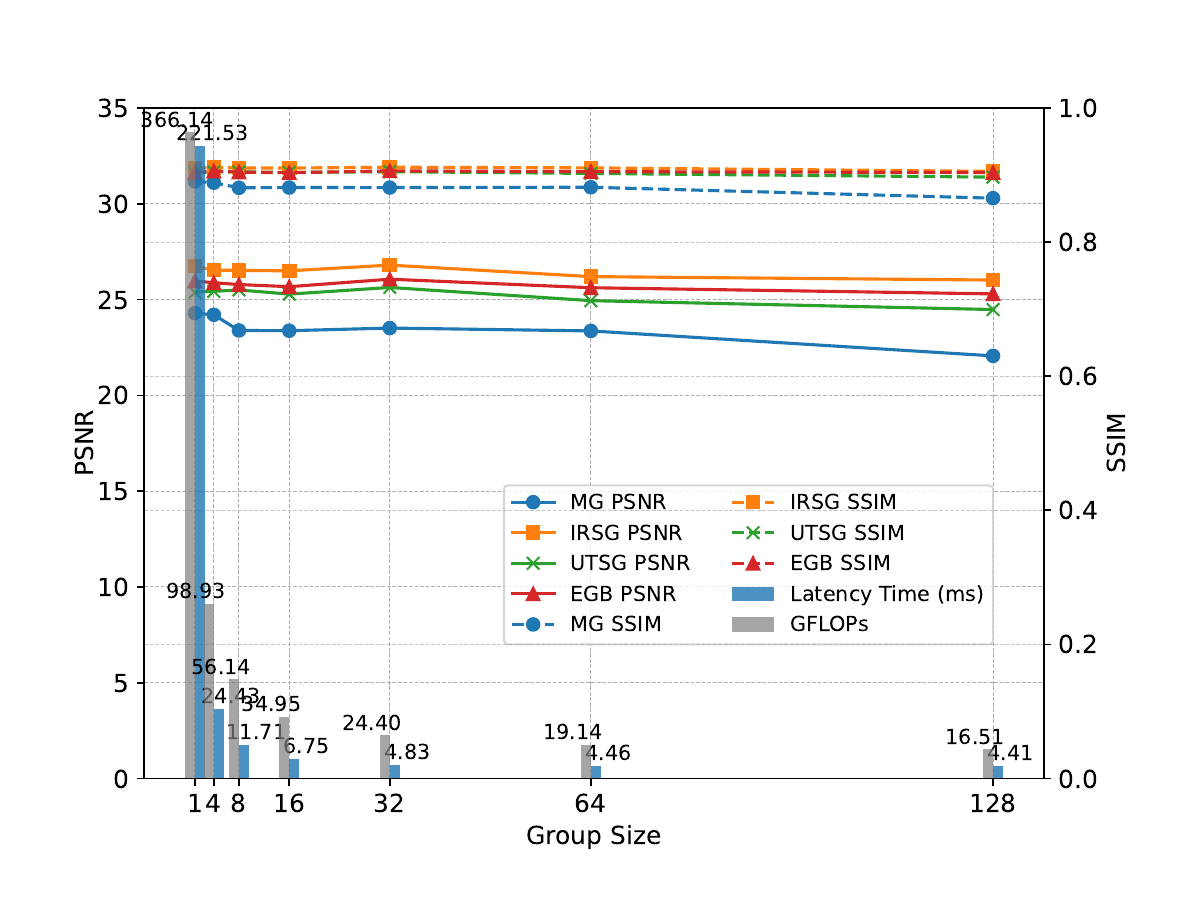}}
\caption{Effects of group size. The experiment compares PSNR and SSIM metrics (lines) against GFLOPs and latency time (bars) across different group sizes for four datasets.}
\label{Granularity}
\end{figure}

As illustrated in Figure \ref{Granularity}, there is a sharp drop in both FLOPs and inference latency as the group size increases from 1 to 16. In contrast, the image quality in SSIM and PSNR is almost constant under most settings. Only the PSNR value of MG dataset decreases slightly. Between a group size of 32 to 128, there is a slight decrease in PSNR and SSIM, implying that increasing group size beyond 32 might compromise the output quality. Moreover, when the group size is set to 1, it is the same as encoding at the granularity of individual points. The results indicate that encoding with group granularity as input to the transformer encoder is significantly more efficient than encoding at the point granularity even when the group size is small. 

\subsection{Ablation Experiment}

\begin{figure}[h]
\centerline{\includegraphics[height=3cm]{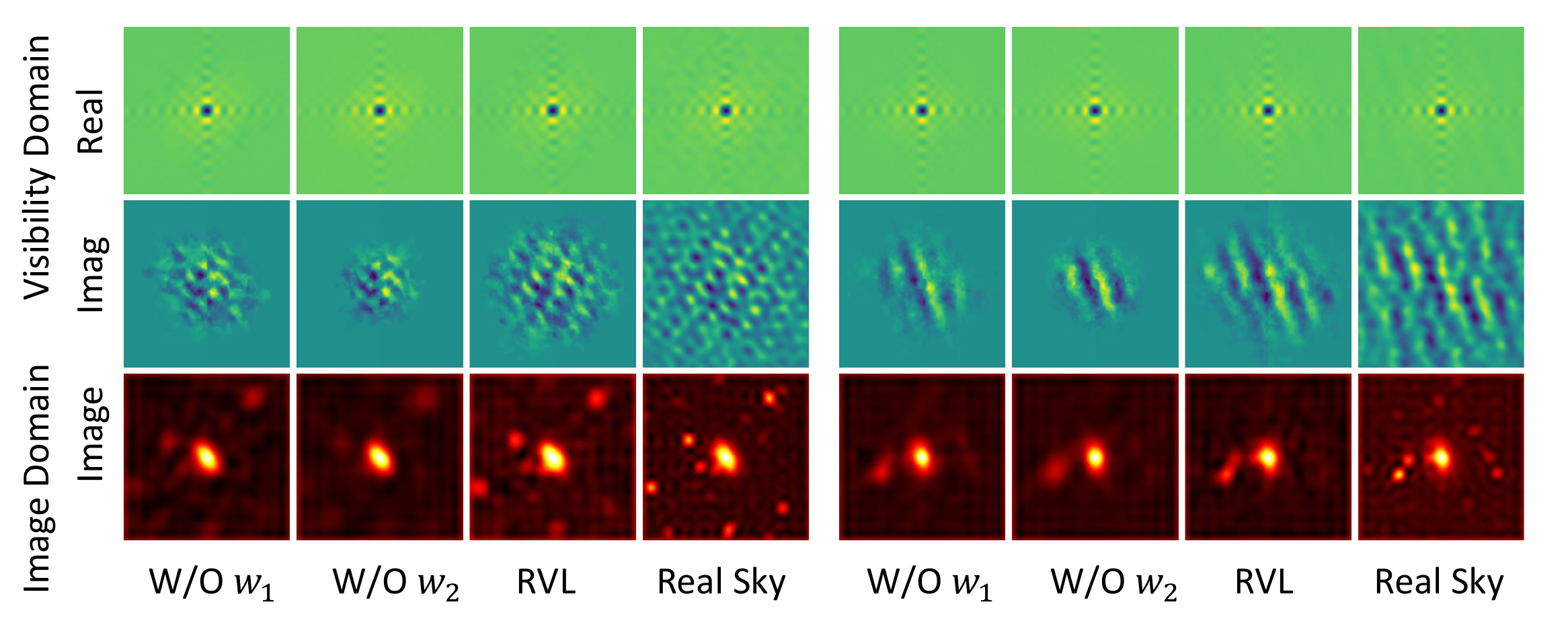}}
\caption{Visual examples of ablation experiments.}
\label{Ablation_img}
\end{figure}

This ablation experiment aims to examine the significance of two weight matrices $w_1, w_2$ in the Radial Visibility Loss by omitting them one at a time. The results in Table \ref{Ablation} and Figure \ref{Ablation_img} show that the full Radial Visibility Loss is the best. Removing either component $w_1$, or $w_2$ results in reduced performance across all metrics.

\begin{table}[h]
    \centering
    \fontsize{6.4}{9}\selectfont
    \caption{Ablation study performance comparison (mean and standard deviation) on different datasets.}
    \label{Ablation}
    \renewcommand{\arraystretch}{1.2} 
    \setlength\tabcolsep{2pt} 
    \begin{tabular}{r|ccc|ccc}
        \hline
        \cline{2-7}
         & \multicolumn{3}{c|}{\textbf{MG}} & \multicolumn{3}{c}{\textbf{IRSG}} \\
         & $LFD\downarrow$ & $PSNR\uparrow$ & $SSIM\uparrow$ & $LFD\downarrow$ & $PSNR\uparrow$ & $SSIM\uparrow$ \\
        \hline
        \textbf{w/o $w_2$} & 1.088 \tiny{(0.309)} & 22.498 \tiny{(2.578)} & 0.884 \tiny{(0.030)} & 0.876 \tiny{(0.305)} & 24.435 \tiny{(2.577)} & 0.904 \tiny{(0.026)} \\
        \textbf{w/o $w_1$} & 1.075 \tiny{(0.317)} & 22.706 \tiny{(2.797)} & 0.882 \tiny{(0.032)} & 0.818 \tiny{(0.283)} & 25.213 \tiny{(2.641)} & 0.907 \tiny{(0.027)} \\
        \rowcolor{gray!20}\textbf{RVL} & \textbf{0.924} \tiny{\textbf{(0.284)}}&\textbf{24.206} \tiny{\textbf{(2.547)}}&\textbf{0.889} \tiny{\textbf{(0.029)}} & \textbf{0.666} \tiny{\textbf{(0.238)}}&\textbf{26.540} \tiny{\textbf{(2.653)}}&\textbf{0.911} \tiny{\textbf{(0.025)}} \\
        \hline
         & \multicolumn{3}{c|}{\textbf{UTSG}} & \multicolumn{3}{c}{\textbf{EGB}} \\
         & $LFD\downarrow$ & $PSNR\uparrow$ & $SSIM\uparrow$ & $LFD\downarrow$ & $PSNR\uparrow$ & $SSIM\uparrow$ \\
        \hline
        \textbf{w/o $w_2$} & 0.908 \tiny{(0.299)} & 23.477 \tiny{(2.574)} & 0.897 \tiny{(0.027)} & 0.933 \tiny{(0.310)} & 23.793 \tiny{(2.507)} & 0.897 \tiny{(0.026)} \\
        \textbf{w/o $w_1$} & 0.854 \tiny{(0.281)} & 24.393 \tiny{(2.627)} & 0.901 \tiny{(0.028)} & 0.864 \tiny{(0.287)} & 24.569 \tiny{(2.500)} & 0.901 \tiny{(0.026)} \\
        \rowcolor{gray!20}\textbf{RVL} & \textbf{0.707} \tiny{\textbf{(0.241)}}&\textbf{25.430} \tiny{\textbf{(2.565)}}&\textbf{0.905} \tiny{\textbf{(0.025)}} & \textbf{0.707} \tiny{\textbf{(0.241)}}&\textbf{25.880} \tiny{\textbf{(2.529)}}&\textbf{0.906} \tiny{\textbf{(0.025)}} \\
        \hline
    \end{tabular}
    \vspace{-1.0em}
\end{table}

\begin{figure}
  \centering
  \begin{subfigure}{1\linewidth}
    \includegraphics[width=8.2cm]{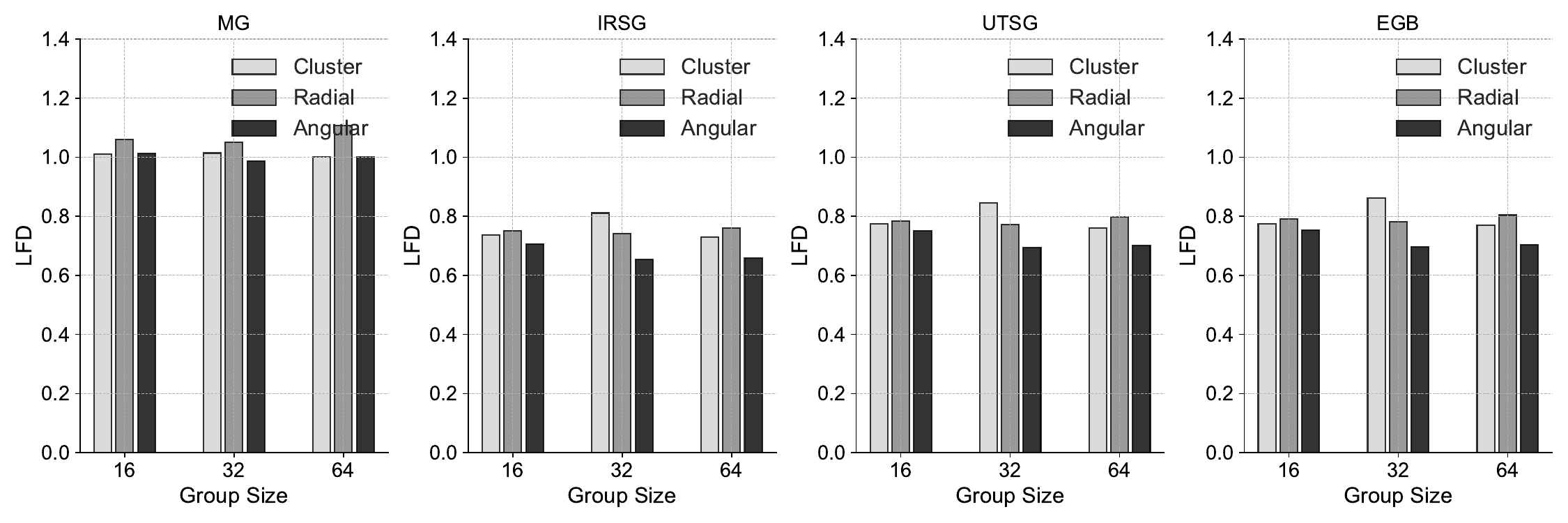} 
  \end{subfigure}
  \\
  \begin{subfigure}{1\linewidth}
    \includegraphics[width=8.2cm]{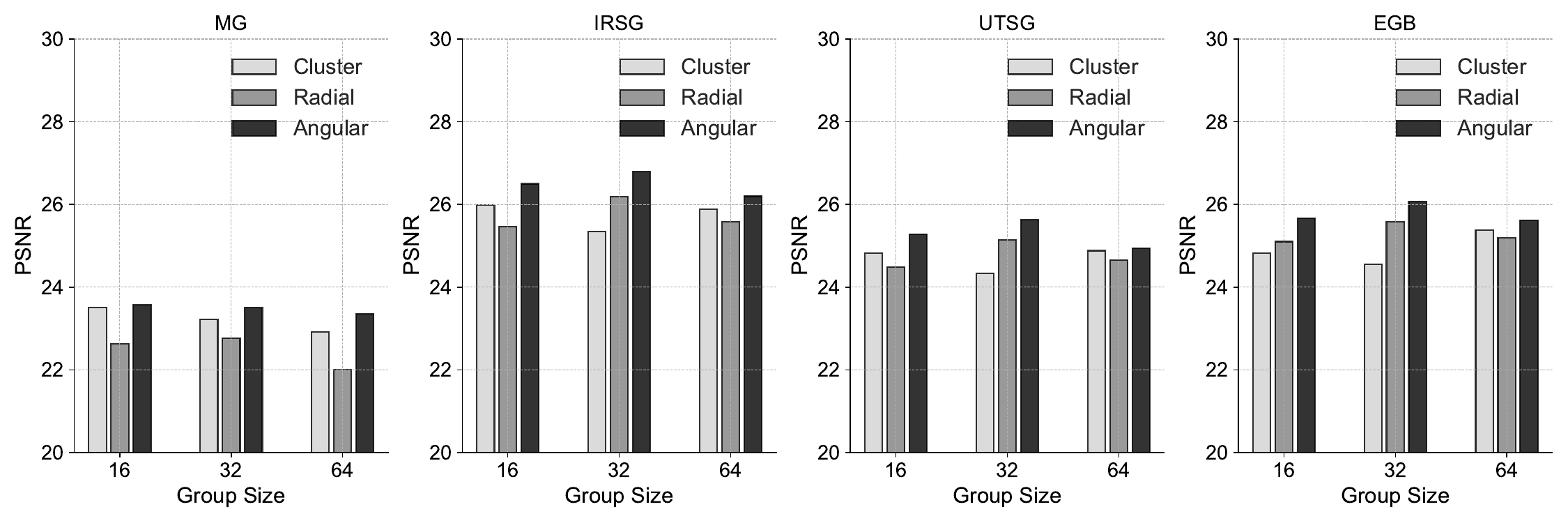}
  \end{subfigure}
  \caption{Comparison of grouping method over various group sizes and different datasets.}
  \label{GroupingMethod}
\end{figure}

\subsection{Effect of grouping method}

We also vary the grouping method and measure its performance impact on PolarRec. We implemented three grouping strategies: (1) grouping by clustering visibility points based on their positions, (2) grouping by the radial coordinate, and (3) grouping by the angular coordinate. Each of these strategies was tested under various group sizes on different datasets. 
The LFD and PSNR results are presented in Figure \ref{GroupingMethod}. Regardless of the group size, grouping the visibility points by angular coordinates always has the best performance.

\section{Conclusion and Future Work}
We have presented PolarRec, reconstructing interferometric visibility with sample points represented in the polar coordinate system. By adopting angular and radial coordinates of visibility points, our method addresses two critical challenges in visibility reconstruction: reconstruction quality and computational efficiency. Our results show that PolarRec markedly improves imaging outcomes while significantly reducing the computation cost, demonstrating a new, feasible, and extendable approach for cross-disciplinary progress in computer vision and radio astronomy.

While PolarRec is designed for radio astronomical data reconstruction, its underlying principles may find utility in other domains requiring sophisticated frequency component analysis and reconstruction, such as magnetic resonance imaging or seismic imaging. Our proposed structural sparse point encoding method may also be used in point cloud data analysis. Future work could explore these potential applications and generalize the work to other research fields.
{
    \small
    \bibliographystyle{ieeenat_fullname}
    \bibliography{main}
}


\end{document}